\documentclass[twocolumn,amsmath,amssymb,10pt,aps]{revtex4-1}
 \usepackage[colorlinks=false]{hyperref}
 \usepackage{lipsum}
 \usepackage{graphicx}
\usepackage{latexsym}
\usepackage{amsmath}
\usepackage{amsthm}
\usepackage{amssymb}
\usepackage{epstopdf} 
\usepackage{enumerate}
\usepackage{setspace}
\usepackage{dcolumn}
\usepackage{bm}
\usepackage{setspace} 
\usepackage{slashed}
\usepackage{color}
\usepackage{youngtab}
\usepackage{tikz}
\usepackage{tikz-3dplot}
\usepackage{braket}
\usepackage{cancel}
\usepackage{subfigure}
\usetikzlibrary{shapes,snakes,arrows,chains,matrix,positioning,scopes,calc}
\usetikzlibrary{decorations.markings}
\tikzset{/pgf/decoration/.cd,
    number of sines/.initial=10,
    angle step/.initial=20,
}
\newdimen\tmpdimen
\pgfdeclaredecoration{complete sines}{initial}
{
    \state{initial}[
        width=+0pt,
        next state=move,
        persistent precomputation={
            \pgfmathparse{\pgfkeysvalueof{/pgf/decoration/angle step}}%
            \let\anglestep=\pgfmathresult%
            \let\currentangle=\pgfmathresult%
            \pgfmathsetlengthmacro{\pointsperanglestep}%
                {(\pgfdecoratedremainingdistance/\pgfkeysvalueof{/pgf/decoration/number of sines})/360*\anglestep}%
        }] {}
    \state{move}[width=+\pointsperanglestep, next state=draw]{
        \pgfpathmoveto{\pgfpointorigin}
    }
    \state{draw}[width=+\pointsperanglestep, switch if less than=1.25*\pointsperanglestep to final, 
        persistent postcomputation={
        \pgfmathparse{mod(\currentangle+\anglestep, 360)}%
        \let\currentangle=\pgfmathresult%
    }]{%
        \pgfmathsin{+\currentangle}%
        \tmpdimen=\pgfdecorationsegmentamplitude%
        \tmpdimen=\pgfmathresult\tmpdimen%
        \divide\tmpdimen by2\relax%
        \pgfpathlineto{\pgfqpoint{0pt}{\tmpdimen}}%
    }
    \state{final}{
        \ifdim\pgfdecoratedremainingdistance>0pt\relax
            \pgfpathlineto{\pgfpointdecoratedpathlast}
        \fi
   }
}

 \usepackage{moreverb}

\begin{document}
\title{  Long-range Coulomb Interaction effects \\on Topological Phase Transitions between Semi-metals and Insulators  }

\author{SangEun Han}
\author{Eun-Gook Moon}
\thanks{egmoon@kaist.ac.kr}

\affiliation{Department of Physics, Korea Advanced Institute of Science and Technology, Daejeon 305-701, Korea}

\date{\today}

\begin{abstract}   
Topological states may be protected by a lattice symmetry in a class of topological semi-metals. 
In three spatial dimensions, the Berry flux around gapless excitations in momentum space defines a chirality concretely, so a protecting symmetry may be referred to as a chiral symmetry.   
 Prime examples include Dirac semi-metal (DSM) in a distorted spinel, BiZnSiO$_4$, protected by a mirror symmetry and DSM in Na$_3$Bi, protected by a rotational symmetry. In these states, topology and a chiral symmetry are intrinsically tied. 
In this work, we investigate characteristics interplay between a chiral symmetry order parameter and instantaneous long-range Coulomb interaction with the standard renormalization group method. 
We show that a topological transition associated with a chiral symmetry is stable under the presence of the Coulomb interaction and the electron velocity always becomes faster than one of a chiral symmetry order parameter. Thus, the transition {\it must not} be relativistic, which implies a supersymmetry is intrinsically forbidden by the long-range Coulomb interaction.  
 {Asymptotically exact} universal ratios of physical quantities such as energy gap ratio are obtained, and connections with experiments and recent theoretical proposals are also discussed.   
 \end{abstract}

\maketitle

The discovery of topological insulators and Weyl / Dirac semi-metals shed new lights on understanding in condensed matter physics \cite{RevModPhys.82.3045,RevModPhys.83.1057,Pesin2010,RevModPhys.90.015001}. 
Topology may enforce non-trivial states such as gapless states or topological orders in either boundary or bulk
unveiling a variety of new phases. 
Almost complete classification has been done for non-interacting systems \cite{kitaev2009periodic,PhysRevB.78.195125,PhysRevB.90.205136}.  
 Interacting topological phases, on the other hand, have intrinsic complexity from the Coulomb interaction.  Even though remarkable advances have been achieved recently \cite{2014arXiv1406.3032M, Wang629,  PhysRevB.91.125147, 2017arXiv170506728W, PhysRevLett.119.227002,2017arXiv171007012M,2017arXiv170509298L}, it is also much desired to have a concrete example whose exact analysis is available with striking Coulomb interaction effects. 
We provide one such example relying on a topological phase transition associated with a chiral symmetry breaking. 

A chiral symmetry is a lattice symmetry which protects energy gapless-ness in a topological semi-metal. 
Namely, an energy gap  (``mass'') is generated by breaking the chiral symmetry. 
In condensed matter, it may be realized by a discrete lattice symmetry in contrast to a continuous symmetry between the left- and right-handed fermions in high energy physics.  
For example, a mirror symmetry is a chiral symmetry in Dirac semi-metals of distorted spinels such as BiZnSiO$_4$ \cite{PhysRevLett.112.036403}, and a translational symmetry is in generic Weyl semi-metals \cite{PhysRevB.83.205101,PhysRevB.85.045124}. 
In Table \ref{tab:materials}, we list examples of topological semi-metals and lattice symmetries. 
The chirality is tied to a topological number in terms of Berry flux (or phase) around gapless excitations {\cite{PhysRevLett.113.046401}}, and thus chiral symmetry breaking can be naturally understood as a topological phase transition.  
 
In this paper, we uncover novel interplay phenomena between a chiral symmetry order parameter, the instantaneous long-range Coulomb interaction, and topological semi-metals. 
We consider a system near a chiral symmetry breaking transition under the long-range Coulomb interaction and adopt the powerful renormalization group (RG) method. 
In three spatial dimensions, we obtain asymptotically exact results because the phase transition is at the upper critical dimension. Remarkably, the chiral symmetry transition is stable under the Coulomb interaction and novel universal ratios which characterize the transition are obtained. 
We emphasize differences between our universal ratios and ones without the Coulomb interaction in the seminal work by Zinn-Justin \cite{ZINNJUSTIN1991105}. 
For example, the Coulomb interaction forces the velocity ratio to be a non-unity, so the emergent ``Lorentz'' symmetry is forbidden.

\begin{table}
\begin{tabular}{|c | c |c | >{$}c<{$} |}
\hline
Phase&Material&Chiral Sym.&N_{f}\\
\hline\hline
DSM&BiZnSiO$_4$ \cite{PhysRevLett.112.036403}&Mirror&1\\
\hline
DSM&Cd$_{3}$As$_{2}$ \cite{Liu2014}, Na$_{3}$Bi \cite{Liu864}&$C_{4}$, $C_{3}$&2\\
\hline
DSM&SrPd$_3$O$_4$ \cite{PhysRevB.95.035102}&$\tilde{C}_{4}$&6\\
\hline
WSM&R$_2$Ir$_2$O$_7$ \cite{PhysRevB.83.205101,PhysRevB.85.045124}&Tran.&4, 12\\
\hline
\end{tabular}
\caption{Examples of topological semi-metals with corresponding chiral symmetries. Reported Dirac / Weyl fermion numbers are listed. DSM (WSM) is for Dirac (Weyl) semi-metal. For WSM, we consider a generic case where Weyl points do not touch a Brillouin zone boundary. In such a case, a translational symmetry (Tran.) plays a role as a chiral symmetry. $C_{n}$ ($\tilde{C}_{n}$) stands for $n$-fold (screw) rotational symmetry. $N_{f}$ is the number of the Dirac (Weyl) points in Brillouin zones.
 } \label{tab:materials}
\end{table}

To illustrate our results, we begin with a kinetic Hamiltonian for topological semi-metals in three spatial dimensions, 
\begin{eqnarray}
H_0=\sum_k \Psi_k^{\dagger}  \mathcal{H}_0(k) \Psi_k= \sum_k \Psi_k^{\dagger} (\varepsilon_a(k)\Gamma^a) \Psi_k, \nonumber
\end{eqnarray}
Generically, three $\Gamma^{a=x,y,z,}$ matrices are necessary, which are $4N_f \times 4N_f$ mutually anti-commuting matrices with a fermion flavor number $N_f$. We introduce a spinor, $\Psi_k$, with $4N_f$ components. Our focus is intrinsic nature of topological semi-metals, so we do not consider a chemical potential term as in  distorted spinnels \cite{PhysRevLett.112.036403}.
The energy functions $\varepsilon_{x,y,z} (k)$ determines properties of topological semi-metals. For example, a nodal line semi-metal may be realized by choosing $\epsilon_z = v_z k_z$, $\epsilon_{x} = \frac{k_x^2 +k_y^2- k_F^2}{2m}$, and $\epsilon_y = \frac{k_z^2}{2M}$ \cite{PhysRevB.90.205136,PhysRevB.95.094502,PhysRevB.93.035138}. Choosing $\epsilon_{a} = v_a k_a$ makes a nodal point semi-metal.  
Our main focus in this paper is point-nodal semi-metals, and we take the simplest case with an isotropic velocity ($v_x=v_y=v_z$) for presentation of our results, $\mathcal{H}_0(k) = v (k_x \Gamma^x + k_y \Gamma^y + k_z \Gamma^z)$ unless stated otherwise (see more general cases in supplementary material \cite{SuppMat}).
We briefly discuss the case of nodal-line semi-metals later. 
 
The chiral symmetry may be realized by introducing an unitary matrix $\Gamma_5 = i \Gamma^x \Gamma^y \Gamma^z$, which commutes with the kinetic Hamiltonian, $[\mathcal{H}_0(k), \Gamma_5]=0$.  One specific representation for $N_f=1$ is $\Gamma^{x,y,z} = \tau^z \otimes \sigma^{x,y,z}$ with Pauli matrices ($\sigma^{x,y,z}, \tau^z$).
The quantum number is assigned by $\Gamma_5$, which defines a Berry flux in topological semi-metals. 
Also, it is easy to show that a matrix $M$ with $\{M,\Gamma^{x,y,z}\} = \{M,\Gamma^5\}=0$ always exists. 
The corresponding operator $\Psi_k^{\dagger} M \Psi_k$ breaks the chiral symmetry, and the anti-commutation relation guarantees that the energy spectrum is gapped by the operator.
  
  To proceed, we consider the action with an imaginary time $\tau = i t$,
\begin{eqnarray}
\mathcal{S} &=&  \int_{x, \tau}  \frac{1}{2} (\nabla \varphi)^2 +\frac{1}{2} (\nabla \phi)^2+ \frac{(  \partial_{\tau} \phi)^2}{2 u^2}   +\frac{r}{2}  \phi^2 + \frac{1}{4!}\frac{\lambda}{u} \phi^4 \nonumber   \\
&+&\int_{x, \tau}  \Psi^{\dagger}(\partial_{\tau} + \mathcal{H}_0 (-i \nabla)) \Psi + i e \varphi \, \Psi^{\dagger} \Psi +g \phi \, \Psi^{\dagger} M \Psi, \nonumber
\end{eqnarray}
with $\int_{x,\tau} = \int d\tau d^3x$.
The chiral symmetry order parameter $\phi$ is introduced, and $g$ characterizes the strength of electron-order parameter coupling. 
Notice that we consider a generic case that the chiral symmetry is a discrete symmetry, so $\phi$ is real. Its generalization to a continuous group is straightforward (SM). 
The electric potential $\varphi$ with an electric charge $e$ captures the instantaneous long-range Coulomb interaction,  $ \frac{e^2}{2}\int d^3x d^3y \frac{n(x) n(y)}{|x-y|}$, which can be obtained by integrating out $\varphi$. 
It is easy to show that coupling between the order parameter and long-range Coulomb interaction is irrelevant because the order parameter couples to the Coulomb interaction non-minimally.  The lowest-order gauge-invariant and symmetric coupling has a form of $\phi^{2}(\nabla\varphi)^{2}$, and such a coupling term is irrelevant in RG analysis.
The boson velocity, $u$, is introduced, which is different from the electron velocity $v$  in general.  
The four dimensionless coupling constants are naturally defined,
 \begin{eqnarray}
 \alpha_e = \frac{e^2}{4\pi v}, \quad \alpha_g =\frac{ g^2}{4\pi v}, \quad y =\frac{u}{v}, \quad \lambda. \nonumber
 \end{eqnarray}
 Notice that one can choose the velocities $v,u$ are always positive, so we only consider the case with $y \ge 0$.
 
Remark that similar actions have been investigated in literatures. For example, without the Yukawa coupling ($g=0$), the actions with symmetry protected semi-metal Hamiltonians have been studied in contexts of non-Fermi liquids and stable double / anisotropic Weyl semi-metals \cite{PhysRevLett.111.206401,PhysRevB.92.045121,PhysRevB.91.235131,Yang2014,Cho2016,PhysRevB.90.075137, PhysRevB.93.241113,PhysRevLett.116.076803}. With a non-zero coupling ($g \neq0$), characteristic phase transitions have been investigated  \cite{PhysRevX.4.041027,PhysRevLett.113.106401,PhysRevB.95.094502} . Furthermore, notorious problems on phase transitions with two-dimensional Fermi-surfaces have been studied in recent literatures, which analyze the actions with similar structures \cite{PhysRevB.95.245109, PhysRevX.7.021010,PhysRevB.94.195135}. 
In this work, we take a generic Hamiltonian of linear Dirac / Weyl semi-metals and find remarkable interplay physics between electrons, order parameters, and the Coulomb interaction in topological nodal semi-metals.  

Stable phases of the action can be easily obtained by taking limits of the coupling constants.  
First, a chiral symmetry broken phase may be realized by a large enough negative $r$, and the mean field analysis gives $\langle \phi \rangle =\sqrt{\frac{-6r}{ \lambda/u}}$ .
Then, the electron and the order parameter energies become gapped, $E_{gap}^e = g |\langle \phi \rangle|$ and $E_{gap}^{\phi} = \sqrt{2 |r| u^2}$. The ratio of the two energy gaps is arbitrary  
\begin{eqnarray}
\mathcal{R}_G \equiv \frac{E_{gap}^{\phi}}{E_{gap}^{e}} = \sqrt{ \frac{y} {12\pi}\frac{\lambda}{\alpha_g}}
\end{eqnarray}
at the mean field level.

Second, a symmetric phase may be realized by a large enough positive $r$. 
The mean-field level analysis gives $\langle \phi \rangle = 0$, and the electron spectrum remains gapless. 
Thus, the total action effectively becomes
\begin{eqnarray}
\mathcal{S}_{sym} = \int_{x, \tau}  \Psi^{\dagger}(\partial_{\tau} + \mathcal{H}_0 (-i \nabla)) \Psi + i e \varphi \, \Psi^{\dagger} \Psi + \frac{1}{2} (\nabla \varphi)^2  \nonumber.
\end{eqnarray} 
The perturbative calculation with the long-range Coulomb interaction gives the electron self-energy, 
\begin{eqnarray}
\Sigma_f(k, i \omega_n) = -  \int_{q, i q_n} \frac{1}{-i\omega_n - i q_n + \mathcal{H}_0(k+q)} \frac{e^2}{q^2}, \nonumber
\end{eqnarray}
and it is easy to show the self-energy is logarithmically divergent. 
The divergence may be cured by the RG analysis where the momentum shell scheme gives 
$
\Sigma_f(k, i \omega_n) \rightarrow - \frac{2 \alpha_e}{3\pi}\log(\frac{\Lambda}{\mu}) \times  \mathcal{H}_0(k)$. $\Lambda$  $(\mu)$ is a UV (IR) scale.  
Furthermore, the self-energy of $\varphi$ is
\begin{eqnarray}
\Pi_e(q) = e^2 \int_{k, i\omega_n}   {\rm Tr} \Big(\frac{1}{-i \omega_n + \mathcal{H}_0(k+q)} \frac{1}{-i \omega_n + \mathcal{H}_0(k)} \Big), \nonumber
\end{eqnarray}
which is also logarithmically divergent, and the momentum shell scheme gives $\Pi_e(q) \rightarrow - \frac{2N_f}{3\pi} \alpha_e \log(\frac{\Lambda}{\mu}) \times q^2 $.

The logarithmic divergences give beta functions of the scale-dependent couplings. 
For example, the velocity beta function is $\frac{d}{d l} v(l) =\frac{2}{3 \pi} \alpha_e(l) v(l)$, with $\Lambda = \mu e^{l}$, and  the fine structure constant, $\alpha_e$, beta function is \cite{PhysRevLett.108.046602}, 
\begin{eqnarray}
\beta_e(l)\equiv  \frac{d}{d l} \alpha_e (l) = - \frac{2}{3\pi} (N_f +1) \alpha_e(l)^2. \label{ebeta}
\end{eqnarray}
The  scale-dependent fine structure constant is obtained by solving the beta function, $\alpha_e(l) = \frac{\alpha_e(0)}{1+\frac{2}{3\pi}(N_f+1) \alpha_e(0) l  }$, 
and in the long distance limit $l \rightarrow \infty$, it becomes $\alpha_e(l) \rightarrow \frac{3 \pi }{2(N_f+1)} \frac{1}{l}$, which manifestly shows the screening of the Coulomb interaction in a long distance limit and the weak-coupling fixed point $(\alpha_e \rightarrow 0)$ is stable. 
 
Before going further, we remark that, in recent interesting works, a new phase is proposed by solving a self-consistent Schwinger-Dyson equation \cite{PhysRevB.92.125115}. 
Ignoring vertex-corrections, a non-Fermi liquid phase is suggested in a strong coupling limit, which cannot be obtained by Eqn.\eqref{ebeta}. 
Higher order corrections should be added, and then a NFL phase can be understood by a stable fixed point at $\alpha_e^* \neq 0$ with $\beta_e(\alpha_e^*) =0$ \cite{PhysRevLett.111.206401,PhysRevLett.113.106401,González2015}. 
All physical quantities receives corrections. For example, the dynamical critical exponent of the NFL is $z = 1- \frac{2}{3\pi} \alpha_e^* +O((\alpha_e^*)^2)$.  

Next, let us consider a quantum phase transition.
The standard mean-field analysis without the Coulomb interaction predicts a second order phase transition at the critical point, $r=r_c$ \cite{ZINNJUSTIN1991105}.
However, it is not obvious whether the Coulomb interaction destabilizes or stabilizes the criticality, a priori. 
An example of the former case is a fluctuation induced first order transition as in superconductors \cite{PhysRevLett.32.292}, 
and an example of the latter case is a magnetic transition in a non-Fermi liquids in pyrochlore systems \cite{PhysRevX.4.041027}. 
Therefore, it is essential to investigate how the Coulomb interaction modifies the nature of the quantum phase transition. 
  
We perform the standard RG analysis in terms of  $\alpha_e, \alpha_g,$ and $\lambda$ and absorb all logarithmic divergences, which are similar to the ones we did for Eqn. \eqref{ebeta}. 
Namely, we find the beta functions by evaluating various Feynman diagrams in SM,  
\begin{eqnarray}
\frac{d }{ d l} \,y \,\,&=& \left( \frac{N_f}{2} (1+y) +\frac{2}{3} \frac{y}{(1+y)^2} \right)y(1-y) \frac{\alpha_g}{\pi} - \frac{2}{3\pi} \alpha_e y \nonumber\\
\frac{d }{d l}\, \lambda \,\, &=& -\frac{3}{16 \pi^2} \lambda^2 + 48 N_f y \alpha^2_g - \frac{1}{2} y^2 \frac{N_f}{\pi}\alpha_g \lambda -\frac{3}{2} \frac{N_f}{\pi} \alpha_g \lambda \nonumber \\
 \frac{d}{dl} \alpha_g&=&-\frac{3 N_f (y+1)^2+2 y (4 y+5)}{3 \pi  (y+1)^2} \alpha_g^2 + \frac{4}{3\pi} \alpha_g \alpha_e \nonumber \\
 \frac{d}{dl} \alpha_e &=& -\frac{2}{3\pi} (N_f +1) \alpha_e^2 + \frac{2}{3\pi} \alpha_e \alpha_g \frac{y (1-y)}{(1+y)^2}. \label{beta}
\end{eqnarray}
Below, we analyze the beta functions step by step. 

{\it1) $\alpha_e=0$ : } First, turn off the electric charge at the microscopic level, $e=0$. This situation may be more naturally realized by a nodal-to-nodeless superconductor transition in topological superconductors \cite{PhysRevLett.113.046401}.    

The beta function of the velocity ratio becomes 
\begin{eqnarray}
\beta_y(l)= \frac{d }{ d l} y &=& \left( \frac{N_f}{2} (1+y) +\frac{2}{3} \frac{y}{(1+y)^2} \right)y(1-y) \frac{\alpha_g}{\pi}.   \nonumber
\end{eqnarray}
Since $\alpha_g$ is positive semi-definite,  it is obvious that $y=1$ is stable at low energy.  
The beta functions of the other two coupling constants become consistent with the previous results \cite{ZINNJUSTIN1991105}
\begin{eqnarray}
\frac{d }{d l}\, \lambda \,\, &=& -\frac{3}{16 \pi^2} \lambda^2 + 48 N_f  \alpha^2_g - 2 \frac{N_f}{\pi} \alpha_g \lambda \nonumber \\
 \frac{d}{dl} \alpha_g&=&- \left(\frac{2 N_f +3}{2 \pi } \right)\alpha_g^2. \nonumber 
\end{eqnarray}
The stable fixed point of the beta functions is $(\lambda, \alpha_g,y) = (0,0,1)$.  
Near the fixed point, 
 we obtain the universal energy gap ratio in the long distance limit ($l \rightarrow \infty$), 
\begin{eqnarray}
\mathcal{R}_G (\alpha_e=0) = \frac{\sqrt{\sqrt{4N_f^2 + 132 N_f +9} +3-2N_f}}{3} . \nonumber
\end{eqnarray}

Few remarks follow. 
The velocity ratio becomes unity at the critical point, which indicates the emergence of the Lorentz symmetry at low energy with a speed of ``light''. 
One can treat the velocity as a constant and be benefited by knowledge of relativistic quantum field theories. 
The criticality is well-known as the Gross-Neveu-Yukawa class, which allows a stable second-order quantum phase transition \cite{sachdev2011,ZINNJUSTIN1991105}. 
The universality class is characterized by universal numbers, for example, the mass gap ratio, $\mathcal{R}_G(\alpha_e=0)$, near the criticality point.

\begin{figure}
\centering
\subfigure®{\includegraphics[height=3.3cm]{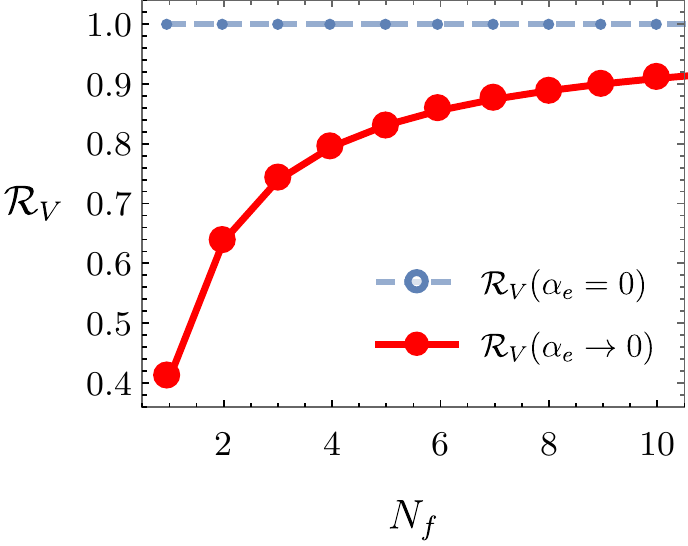}}\hfill
\subfigure{\includegraphics[height=3.3cm]{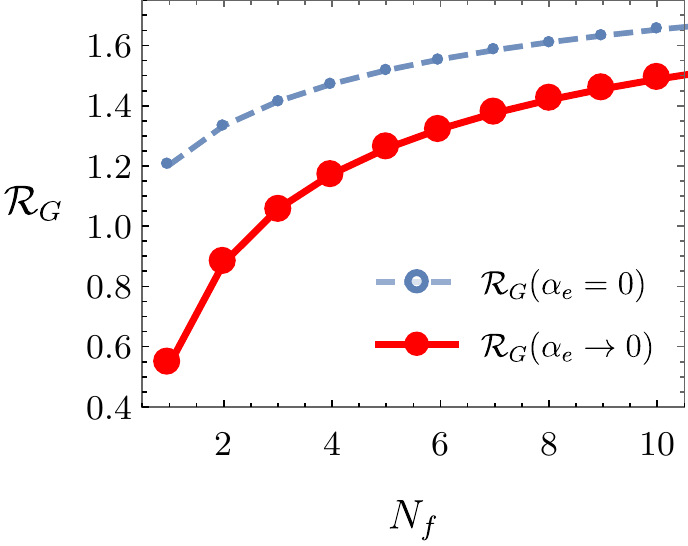} }
\caption{ Universal ratios of velocity, $\mathcal{R}_{V}(\alpha_{e})$, and mass (energy) gap, $\mathcal{R}_G(\alpha_e)$.  The ratios $\mathcal{R}_{V,G}(\alpha_e\rightarrow0)$ (red plane line) are obtained  in this work. For comparison, we illustrate the ratios with the Lorentz invariance ($\mathcal{R}_{V,G}(\alpha_e=0)$) \cite{ZINNJUSTIN1991105} (blue dotted line).  In the limit of $N_{f}\rightarrow\infty$, the Coulomb interaction is fully screened and the relativistic behavior is recovered. }\label{ratio}
\end{figure}

{\it2) $\alpha_e \rightarrow 0$  : }
With the electric charge $e \neq0$, intricate and intriguing interplay physics appears. 
Happily, our RG calculation shows only logarithmic divergences and is fully controlled. 
Full analysis of the beta functions, Eqn. (3), gives a stable fixed point $(\lambda, \alpha_g, y, \alpha_e) = (0,0, y_0,0)$. 
The velocity ratio $y_0$ is a non-unity number which depends on $N_f$. 
Fig.\ref{ratio} clearly shows differences between universal ratios with and without the long-range Coulomb interaction. 
For example, with $N_f=1$, the ratio of the velocities, $\mathcal{R}_V \equiv \frac{u}{v} $, is $\mathcal{R}_V(\alpha_e \rightarrow 0) \rightarrow  \frac{1}{\sqrt{6}} $ in contrast to $\mathcal{R}_V(\alpha_e=0)  \rightarrow 1$.
Thus, taking the limit $l \rightarrow \infty$ and the zero limit of $\alpha_e $ does not commute. 

Few remarks follow. 
First, the presence of the stable fixed point indicates the quantum phase transition is stable under the long-range Coulomb interaction. 
This is distinctly different from the fluctuation induced first-order phase transitions in s-wave superconductors \cite{PhysRevLett.32.292}, which is similar to the Coleman-Weinberg theory \cite{PhysRevD.7.1888}. 
We notice that the order parameter $\phi$ in our system is electric charge-neutral, so there is no direct coupling between $\phi$ and $\varphi$.

Second, the electron and order parameter can be distinguished by their energy spectrums with different velocities. 
Our numerical calculation shows the velocity ratio is monotonically increasing with $N_f$, and we analytically find $y_0 = 1- \frac{1}{N_f}+O(\frac{1}{N_f^2})$ for $N_f \gg1$. Thus, the Lorentz symmetry, which guarantees all velocities are the same, is forbidden by the long-range Coulomb interaction. 
One important consequence of the forbidden Lorentz symmetry is that supersymmetry which exchange fermions and bosons is intrinsically forbidden. 
Our calculation implies that the Lorentz symmetry and associated supersymmetry can be hosted by topological superconductors with the Meissner effect \cite{Grover1248253,PhysRevLett.116.100402,PhysRevLett.119.107202,sungsiklee-susy,PhysRevLett.114.237001,PhysRevLett.118.166802}.  
Notice that we focus on the Hilbert space of electrons with the instantaneous long-range Coulomb interaction, which is different from electrons with $U(1)$ gauge fields \cite{PhysRevB.86.165127,PhysRevB.87.205138, Roy2016}. 

Third, our analysis is exact in a sense that the scaling limit ($l\rightarrow \infty$) gives only logarithmic corrections in three spatial dimensions, which is similar to some features of quasi-local strange metals \cite{PhysRevB.91.125136,doi:10.1146/annurev-conmatphys-031016-025531}.  
One next intriguing question is the stability analysis under the long-range Coulomb interaction in two spatial dimensions (2d). 
In 2d, fluctuations are more relevant, but the Coulomb interaction is still the inverse-distance potential (marginal). 
In principle, two scenarios are possible. First, the Coulomb interaction becomes irrelevant, and the Lorentz symmetry would emerge at a chiral symmetry transition \cite{PhysRevB.80.075432}. Second, similar to our analysis in 3d, the Coulomb interaction would forbid the Lorentz symmetry and associated supersymmetry. 
We expect that similar interplay physics appears in nodal-line topological semimetals as in previous literature \cite{PhysRevB.93.035138,PhysRevB.95.094502} (see also SM).

\begin{figure}
\centering
\includegraphics[width=\linewidth]{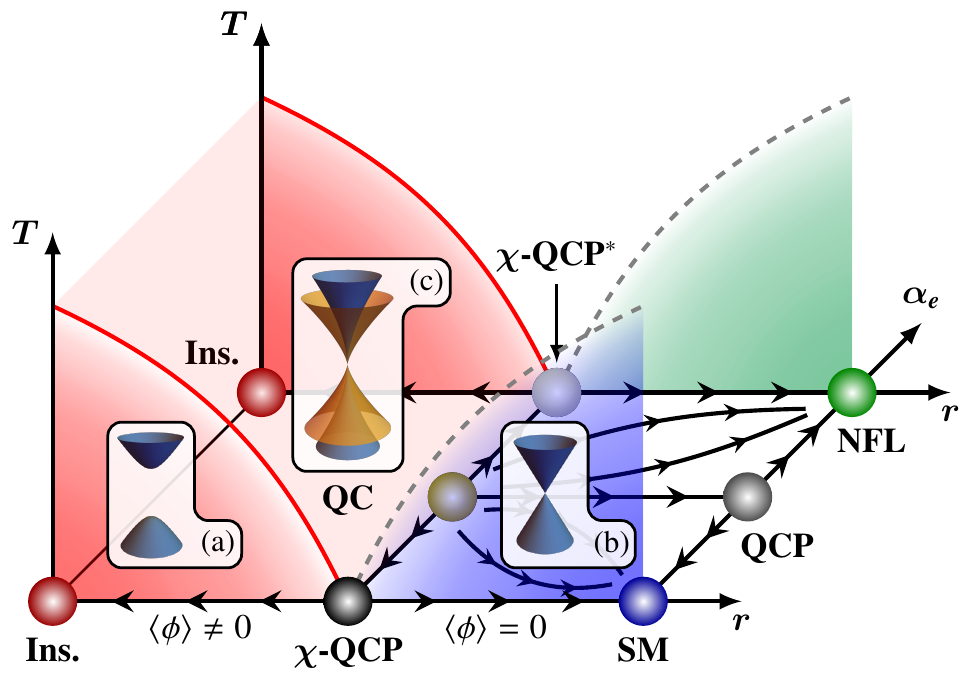}
\caption{Phase diagram of chiral symmetry breaking Transitions with schematic RG flows. SM, Ins., NFL, and QC are for semi-metal, insulator, non-fermi liquid, and quantum critical region, respectively. $\braket{\phi}\neq0$ is for the chiral symmetry-broken phase. $\chi$-QCP, $\chi$-QCP$^{*}$, and QCP are for the quantum critical points between the phases.  (a), (b), and (c) represent the dispersion relations for Ins, SM, and QC, respectively. }\label{phase}
\end{figure}

{\it3) $\alpha_e^* \neq 0$  : }
Our beta function analysis can be further applied to the chiral symmetry breaking out of the NFL phase with a reasonable assumption. 
As mentioned above, the NFL phase can be understood by a fixed point with a non-zero finite coupling constant $\alpha_e^*$. 
The beta functions with $\alpha_e^*$ can be analyzed by using $1/N_f$ expansion with the number of flavors $N_f$ if $\alpha_e^*$ is not too large. 
For example, the chiral order parameter coupling constant beta function becomes 
\begin{eqnarray}
 \frac{d}{dl} \alpha_g&=&-\frac{N_f }{\pi } \alpha_g^2 + \frac{4}{3\pi} \alpha_g \alpha_e^*. \nonumber 
\end{eqnarray}
We find a stable fixed point, $(\lambda^* =0, \alpha_g^*=\frac{4 \alpha_e^*}{3N_f}, y^* = 0)$.
It is obvious that the fixed point is controlled by $1/N_f$, which allows us to analyze the fixed point concretely. 
Again, the existence of the stable fixed point indicates that the chiral symmetry breaking out of NFL is stable. 

Notice that the zero velocity ratio ($y^*=0$) indicates that the Lorentz invariance is also severely broken. 
Our evaluation of dynamical critical exponents of order parameters and electrons gives $z_{e} =z_{\phi}=1 -\frac{2\alpha_{e}^{*}}{3\pi} + O((\alpha_e^*)^2))$. 
In spite of the same dynamical critical exponents, higher order corrections induce the zero velocity ratio. 
This fact indicates that the electron, order parameter, and the Coulomb interaction are strongly correlated, and the sub-linear dynamical critical exponents stabilize the chiral symmetry breaking transition. 

In FIG. \ref{phase}, we summarize our analysis in terms of phase diagrams with schematic RG flows. The RG flow is asymptotically exact near $\alpha_e =0$. 
And for a non-Fermi liquid phase (large enough $\alpha_e$), the RG flow is controlled by  $1/N_f$.

With the above results, we discuss the implication of our results to experiments in topological semi-metals. 
As suggested by recent first principle calculations, various oxides materials are proposed to realized topological semi-metals  including  BiZnSiO$_4$ \cite{PhysRevLett.112.036403} and SrPd$_3$O$_4$ \cite{PhysRevB.95.035102}. 
We expect they are perfect venues to observe our results. 
In addition to the calculated universal ratios, the logarithmic and power corrections to physical quantities are smoking gun signals. 
For example, the velocity renormalization similar to one in graphene \cite{PhysRevLett.99.226803} can be obtained by the beta function. Setting temperature as a scale variable, the velocity becomes $v(T_{0})\left( \log\left(\frac{T_{0}}{T}\right) \right)$ with a given temperature $T_0$. Especially, electronic contribution to the specific heat becomes 
$
C(T) \sim \frac{T^3}{\left(\log\left(\frac{T_{0}}{T}\right)\right)^3},
$ and remarkably the order parameter contributions are automatically determined by the velocity ratio $y_0$ if the systems are not in NFL. 
In the NFL phase, the velocity is renormalized with the power $z_{e}-1$, and the specific heat would shows temperature dependence $C(T) \sim T^{3+\frac{2 \alpha_e^*}{\pi}} $. 

In conclusion, we investigate chiral symmetry breaking with long-range Coulomb interaction in topological semi-metals and show that the transition is stable. 
We obtain {\it exact} universal ratios of physical quantities such as energy gap and velocity and find the Lorentz symmetry is forbidden in 3d topological semi-metals under the presence of the long-range Coulomb interaction. We also suggest experimental signatures of our quantum criticality.

\begin{acknowledgments}
We thank  G. Chen and G. Cho for useful discussions and comments. This work was supported by the POSCO Science Fellowship of POSCO TJ Park Foundation and NRF of Korea under Grant No. 2017R1C1B2009176.\\
\end{acknowledgments}

\onecolumngrid
\clearpage
\begin{center}
\textbf{\large Supplementary materials for ``Long-range Coulomb Interaction effects on Topological Phase Transitions between Semi-metals and Insulators''}
\end{center}
\begin{center}
{SangEun Han and Eun-Gook Moon}\\
\emph{Department of Physics, Korea Advanced Institute of Science and Technology, Daejeon 305-701, Korea}
\end{center}
\setcounter{equation}{0}
\setcounter{figure}{0}
\setcounter{table}{0}
\setcounter{page}{1}

\appendix

\section{Notations and Feynmann diagrams }
We provide more information about notations and Feynman diagrams for Dirac / Weyl semi-metals with a chiral order parameter and long range Coulomb interaction.  
The total action is 
\begin{align*}
\mathcal{S}=&\int d^3x d \tau\sum_{a=1}^{N_{f}}\left[ \psi_{a}^{\dagger}(\partial_{\tau}+\mathcal{H}(-i\nabla))\psi_{a}+ie\varphi(\psi_{a}^{\dagger}\psi_{a})+g\phi(\psi_{a}^{\dagger}M\psi_{a})\right]\\
&+\int d^3x d \tau\left[\frac{1}{2}(\nabla\varphi)^{2}+\frac{1}{2}\left(\frac{(\partial_{\tau}\phi)^{2}}{u}+(\nabla\phi)^{2}\right)+\frac{1}{4!}\frac{\lambda}{u}\phi^{4}\right]
\end{align*}
where $\mathcal{H}_{0}(k)=v\sum_{a=x,y,z}k_{a}\Gamma^{a}$ which satisfies $\{\Gamma_{i},\Gamma_{j}\}=2I\delta_{ij}$.
The propagators of the electron, order parameter, and instantaneous Coulomb interaction are
\begin{align*}
G_{f}(\bm{k},\omega)=&\frac{1}{-i\omega+\mathcal{H}_{0}(k)},\\
G_{c}(\bm{k},\omega)=&\frac{1}{\bm{k}^{2}},\\
G_{o}(\bm{k},\omega)=&\frac{1}{\omega^{2}/u^{2}+k^{2}},
\end{align*}

\begin{figure}[h]
\centering
\subfigure[]{
\includegraphics{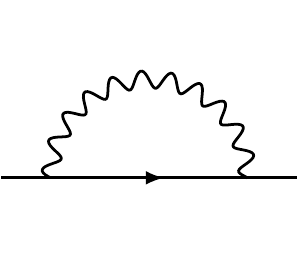}
\label{fig1:a}}
\subfigure[]{
\includegraphics{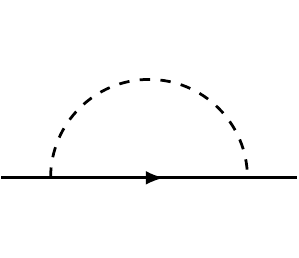}
\label{fig1:b}}
\subfigure[]{
\includegraphics{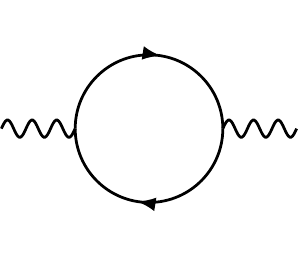}
\label{fig1:c}}
\subfigure[]{
\includegraphics{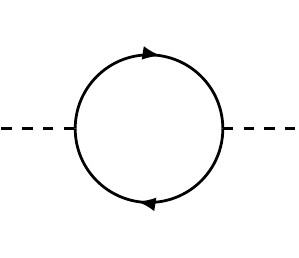}
\label{fig1:d}}

\subfigure[]{
\includegraphics{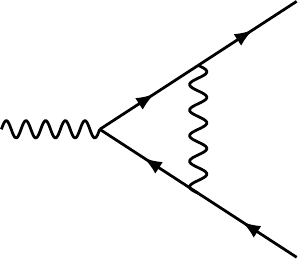}
\label{fig1:e}}
\subfigure[]{
\includegraphics{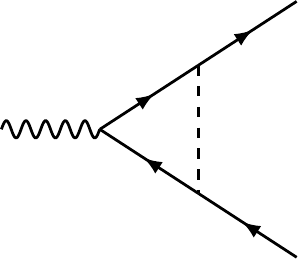}
\label{fig1:f}}
\subfigure[]{
\includegraphics{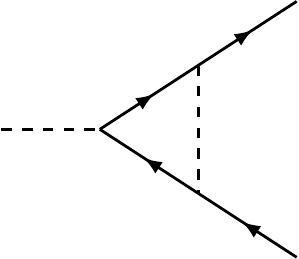}
\label{fig1:g}}
\subfigure[]{
\includegraphics{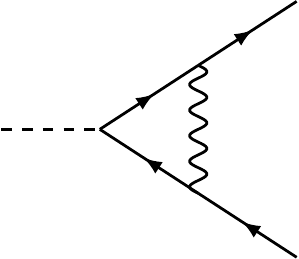}
\label{fig1:h}}

\subfigure[]{
\includegraphics{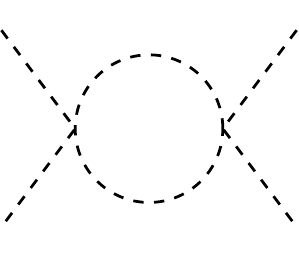}
\label{fig1:i}}
\subfigure[]{
\includegraphics{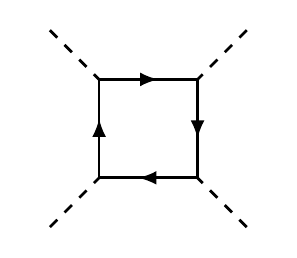}
\label{fig1:j}}
\caption{The one-loop leading order corrections for the self energies and the vertexes. The line with arrowhead, dashed line, and wavy line stand for the fermion, the order parameter, and the Coulomb interaction, respectively.}\label{fig:feynmann}
\end{figure}

First of all, the fermion self energy is given by Fig.\ref{fig1:a} and Fig.\ref{fig1:b} at the one-loop order
\begin{align*}
\Sigma_{f}(\omega,\bm{k})=&g_{b}^{2}\int_{\Omega,q}M_{b}G_{f}(\omega+\Omega,\bm{k}+\bm{q})M_{b}G_{b}(\Omega,\bm{q})
,
\end{align*}
where $\int_{\Omega,q}=\int dq^{3}/(2\pi)^{3}(\int^{-\Lambda e^{\text{-}\ell}}_{-\Lambda}+\int^{\Lambda}_{\Lambda e^{\text{-}\ell}})d\Omega/2\pi$ is momentum-frequency integration, in $g_{b}$ and $M_{b}$, $b=c,o$ where $c$ is for the Coulomb interaction, $g_{c}=ie$, $M_{c}=I$, and $o$ is for the chiral symmetry breaking order parameter interaction, $g_{o}=g$, $M_{o}=M$ which satisfies $\{M,\mathcal{H}_{0}\}=0$.\\
The boson self energy by fermion loop is given by Fig.\ref{fig1:c} (for the Coulomb interaction) or Fig.\ref{fig1:d} (for the order parameter),
\begin{align*}
\Pi_{b}(\omega,\bm{k})=&-g_{b}^{2}\int_{\Omega,q}\text{Tr}\left[ M_{b}G_{f}(\Omega-\omega/2,\bm{q}-\bm{k}/2)M_{b}G_{f}(\Omega+\omega/2,\bm{q}+\bm{k}/2) \right].
\end{align*}
Note that from the fermion and order parameter self-energies, we can obtain the flow equation of the velocity ratio $y\equiv u/v$, $dy/d\ell$. $dy/d\ell$ does not contain the term which is not proportional to $y$. One way to see this is that the divergences of the fermion and boson self-energy are linear or quadratic, respectively. The self-energy divergences become logarithmic after taking derivatives with momentum and momentum square, respectively. Thus, the correction terms are proportional to the bare terms which are another good features in $d=3$, and one can show they are all proportional to $y$. When $z=2$, the case that contains the term which is not proportional to the bare term  is reported \cite{PhysRevB.93.205138}.\\
The correction for vertex between fermion and boson is given by Fig.\ref{fig1:e}, \ref{fig1:f}, \ref{fig1:g}, and Fig.\ref{fig1:h},
\begin{align*}
\delta\Gamma_{bb'}=&g_{b'}^{2}\int_{\Omega,q}M_{b'}G_{f}(\Omega,\bm{q})M_{b}G_{f}(\Omega,\bm{q})M_{b'}G_{b}(\Omega,\bm{q}),
\end{align*}
where $\delta\Gamma_{bb'}$ means the vertex correction for $g_{b}\phi_{b}(\Psi^{\dagger}M_{b}\Psi)$ by $g_{b'}\phi_{b'}(\Psi^{\dagger}M_{b'}\Psi)$ ($\phi_{c}=\varphi$ and $\phi_{o}=\phi$). For example, $\delta\Gamma _{oo}$ and $\delta\Gamma_{oc}$ are
\begin{align*}
\delta\Gamma_{oo}=&g^{2}\int_{\Omega,q}MG_{f}(\Omega,q)MG_{f}(\Omega,q)MG_{o}(\Omega,q)\\
=&-g^{2}M\int\frac{dq^{3}}{(2\pi)^{3}}2\int^{\Lambda}_{\mu\equiv\Lambda e^{-\ell}}\frac{d\Omega}{2\pi}\frac{1}{\Omega^{2}+v^{2}q^{2}}\frac{1}{\Omega^{2}/u^{2}+q^{2}}\\
=&
-\frac{\alpha_{g}}{\pi}\frac{y}{1+y}\ell M,\\
\delta\Gamma_{oc}=&(ie)^{2}\int_{\Omega,q}G_{f}(\Omega,q)MG_{f}(\Omega,q)G_{c}(\Omega,q)\\
=&
\frac{\alpha_{e}}{\pi}\ell M,
\end{align*}
where $y=u/v$ and $\ell=\log\Lambda/\mu$.\\
The $\phi^{4}$ vertex correction by $\phi^{4}$ vertex is given by Fig.\ref{fig1:i},
\begin{align*}
\delta\lambda_{\lambda}=&\frac{3}{2}\left(\frac{\lambda}{u}\right)^{2}\int_{\Omega,q}G_{o}(\Omega,\bm{q})G_{o}(\Omega,\bm{q})\\
=&
\frac{\lambda}{u}\frac{3\lambda}{16\pi^{2}}\ell.
\end{align*}
The $\phi^{4}$ vertex correction by the fermion loop is given by Fig.\ref{fig1:j},
\begin{align*}
\delta\lambda_{g}=&-6g^{4}N_{f}\int_{\Omega,q}\text{Tr}\left[ G_{f}(\Omega,\bm{q})MG_{f}(\Omega,\bm{q})MG_{f}(\Omega,\bm{q})MG_{f}(\Omega,\bm{q})M \right]\\
=&
-\frac{\lambda}{u}\frac{48N_{f}y\alpha_{g}^{2}}{\lambda}\ell.
\end{align*}
Note that this generates the term which is not proportional to $\lambda$ in $d\lambda/d\ell$.

\section{Universal ratios}
In this section, we discuss the universal ratios, $\mathcal{R}_{G}$, $\mathcal{R}_{\alpha}$, and $\mathcal{R}_{V}$ defined as
\begin{align*}
\mathcal{R}_{G}\equiv&\sqrt{\frac{y}{12\pi}\frac{\lambda}{\alpha_{g}}},\\
\mathcal{R}_{\alpha}\equiv&\frac{\alpha_{e}}{\alpha_{g}},\\
\mathcal{R}_{V}\equiv&\frac{u}{v}=y.
\end{align*}
To obtain these universal ratios, we need the asymptotic behaviors of $y$, $\lambda$, $\alpha_{g}$, and $\alpha_{e}$ near the fixed point. Therefore, we will find those for $\alpha_{e}=0$, $\alpha_{e}\neq0$, and $\alpha_{e}^{*}\neq0$, and obtain the universal ratios $\mathcal{R}_{G}$, $\mathcal{R}_{\alpha}$, and $\mathcal{R}_{V}$. All the results are summarized at Fig.\ref{fig:univ_ratio}.

\subsection{Without Coulomb interaction ($\alpha_{e}=0$)}
First, let us consider $\alpha_{e}=0$ case. When $\ell\rightarrow\infty$, $\lambda$ and $\alpha_{g}$ will be $\lambda\rightarrow C_{\lambda}/\ell$ and $\alpha_{g}\rightarrow C_{\alpha_{g}}/{\ell}$. With these ans\"{a}tze, solving the flow equation, we have
\begin{align*}
C_{\alpha_{g}}(N_{f},y)=&\frac{3\pi(1+y)^2}{3N_{f}(1+y)^{2}+2y(5+4y)},\\
C_{\lambda}(N_{f},y)=&\frac{4\pi^{2}}{9N_{f}(1+y)^{2}+6y(5+4y)}\\
&\times\left( 20y+16y^{2}-3N_{f}(1+y)^{2}(1+y^{2})\right.\\
&\left.+\sqrt{16y^{2}(5+4y)^{2}+9N_{f}^{2}(1+y)^{2}(1+y^{2})^{2}-24N_{f}y(1+y)^{2}(-49-104y-49y^{2}+4y^{3})} \right).
\end{align*}
Regardless of $N_{f}$, $y\rightarrow1$ when $\ell\rightarrow\infty$. Then
\begin{align*}
C_{\alpha_{g}}(N_{f})=&\frac{2\pi}{3+2N_{f}},\\
C_{\lambda}(N_{f})=&\frac{8\pi^{2}}{3(3+2N_{f})}(3-2N_{f}+\sqrt{9+132N_{f}+4N_{f}^{2}}).
\end{align*}
Thus, when $\ell\rightarrow\infty$ with $\alpha_{e}=0$, $\mathcal{R}_{G}$ is
\begin{align*}
\mathcal{R}_{G}(\alpha_{e}=0)=&\sqrt{\frac{y}{12\pi}\frac{\lambda}{\alpha_{g}}}\\
\rightarrow&\sqrt{\frac{1}{12\pi}\frac{C_{\lambda}}{C_{\alpha_{g}}}}\\
=&\frac{1}{3}\sqrt{3-2N_{f}+\sqrt{9+132N_{f}+4N_{f}^{2}}}.
\end{align*}
For $N_{f}=1$, $\mathcal{R}_{G}=\sqrt{1+\sqrt{145}}/3\approx1.204$.\\
Since $\alpha_{e}=0$, $\mathcal{R}_{\alpha}=0$ for all $N_{f}$.
\subsection{With Coulomb interaction ($\alpha_{e}\rightarrow0$)}
When we have the Coulomb interaction, we can obtain $\mathcal{R}_{G}$, $\mathcal{R}_{\alpha}$, and $\mathcal{R}_{V}$ in the same way as before,\\
\begin{align*}
\mathcal{R}_{G}=&\frac{1}{2}\sqrt{\frac{y}{3\pi}\frac{C_{\lambda}(N_{f},y)}{C_{\alpha_{g}}(N_{f},y)}},\\
\mathcal{R}_{\alpha}=&\frac{3(N_{f}(1+y)^{2}+2y(2+y))}{2(3+N_{f})(1+y)^{2}},\\
\mathcal{R}_{V}=&y=-\frac{1}{2}+\frac{1}{6}\sqrt{9-\frac{16}{N_{f}}-\frac{4}{3+N_{f}}-\frac{8(144-42N_{f}+25N_{f}^{2}+18N_{f}^{3})+2^{1/3}a_{1}(N_{f})^{2/3}}{2^{2/3}N_{f}(3+N_{f})a_{1}(N_{f})^{1/3}}}\\
&+\frac{1}{6}\left(18-\frac{32}{N_{f}}-\frac{8}{2+N_{f}}+\frac{8(144-42N_{f}+25N_{f}^{2}+18N_{f}^{3})+2^{1/3}a_{1}(N_{f})^{2/3}}{2^{2/3}N_{f}(3+N_{f})a_{1}(N_{f})^{1/3}}\right.\\
&+\left. \frac{2^{1/6}\times18(12+11N_{f}+3N_{f}^{2})}{\sqrt{N_{f}(3+N_{f})(2^{2/3}(-48+7N_{f}+9N_{f}^{2})-2^{1/3}a_{1}(N_{f})^{1/3}-8(144-42N_{f}+25N_{f}^{2}+18N_{f}^{3})/a_{1}(N_{f})^{1/3})}  }\right)^{1/2},
\end{align*}
where 
\begin{align*}
a_{1}(N_{f})\equiv
&16(b_{1}(N_{f})+9c_{1}(N_{f})^{1/2}),\\
b_{1}(N_{f})=&-1728-1431N_{f}-1143N_{f}^{2}-4580N_{f}^{3}-2889N_{f}^{4}-486N_{f}^{5},\\
c_{1}(N_{f})=&93312N_{f}+45441N_{f}^{2}+234090N_{f}^{3}+304317N_{f}^{4}+247068N_{f}^{5}\\
&+357153N_{f}^{6}+340510N_{f}^{7}+157701N_{f}^{8}+34596N_{f}^{9}+2916N_{f}^{10},\\
C_{\lambda}(N_{f},y)=&\frac{4\pi^{2}}{9N_{f}^{2}(1+y)^{2}+18y(1+2y)+N_{f}(9+48y+33y^{2})}\\
&\times\left( 12y(1+2y)-3N_{f}^{2}(1+y)^{2}(1+y^{2})-N_{f}(21+22y+14y^{2}+18y^{3}+9y^{4})+\right.\\
&\left.+\sqrt{1296N_{f}(3+N_{f})^{2}y(1+y)^{4}+(-12y(1+2y)+3N_{f}^{2}(1+y)^{2}(1+y^{2})+N_{f}(21+22y+14y^{2}+18y^{3}+9y^{4}))^{2}} \right),\\
C_{\alpha_{g}}(N_{f},y)=&\frac{3\pi(3+N_{f})(1+y)^{2}}{3N_{f}^{2}(1+y)^{2}+6y(1+2y)+N_{f}(3+16y+11y^{2})}.
\end{align*}
For example, when $N_{f}=1$, $\mathcal{R}_{V}(\alpha_{e}\rightarrow0)=y=1/\sqrt{6}$, $\mathcal{R}_{G}(\alpha_{e}\rightarrow0)=0.541$, and $\mathcal{R}_{\alpha}(\alpha_{e}\rightarrow0)=\tfrac{9}{200}(8\sqrt{6}-3)$.

\subsection{With finite Coulomb interaction ($\alpha_{e}^{*}\neq0$)}\
Let us assume that $\alpha_{e}^{*}=\text{const.}\neq0$ when $\ell\rightarrow\infty$. In this case, $\alpha_{g}^{*}$ become
\begin{align*}
\alpha_{g}^{*}=\frac{4(1+y)^{2}}{3N_{f}(1+y)^{2}+2y(5+4y)}\alpha_{e}^{*}.
\end{align*}
When $N_{f}\rightarrow\infty$, $\alpha_{g}^{*}\rightarrow\frac{4}{3N_{f}}\alpha_{e}^{*}$. For $\alpha_{g}^{*}\neq0$, $y\rightarrow0$ as $\ell\rightarrow\infty$. For $\lambda$,
\begin{align*}
\frac{d\lambda}{d\ell}=&-\lambda\left(\frac{3\lambda}{16\pi^{2}}+\frac{3\alpha_{g}^{*}}{2\pi}N_{f} \right)
\end{align*}
So, $\lambda\rightarrow0$ as $\ell\rightarrow\infty$.\\
Therefore, when $\alpha_{e}^{*}\neq0$, then $y,\lambda\rightarrow0$, so $\mathcal{R}_{G}=0$ for all $N_{f}$. And
\begin{align*}
\mathcal{R}_{e}=\frac{\alpha_{e}}{\alpha_{g}}=\frac{3}{4}N_{f}.
\end{align*}

\subsection{Relativistic regime}
\begin{figure}[h]
\centering
\includegraphics[width=0.322\linewidth]{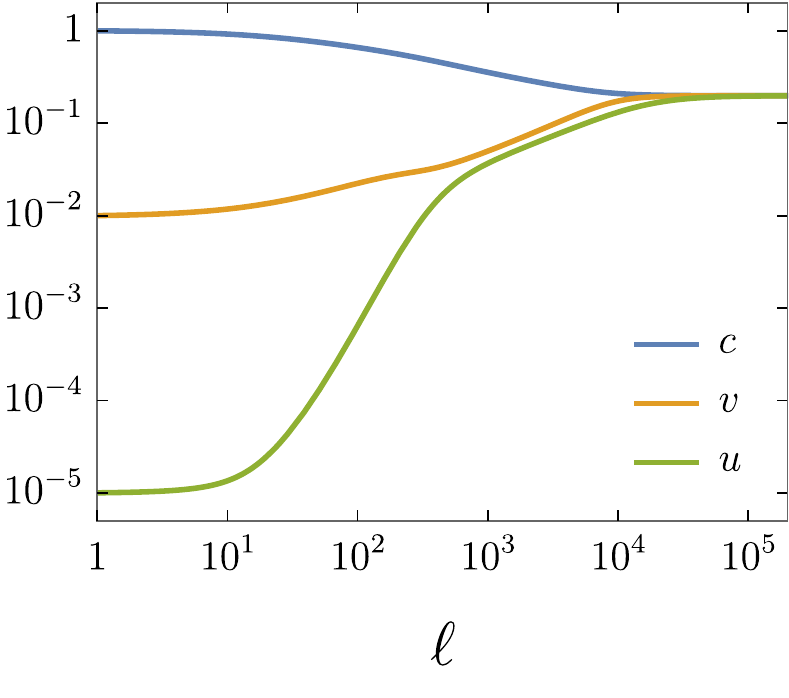}

\caption{The flows of $c$, $v$, and $u$ for the initial values $c_{0}=1$, $v_{0}=10^{-2}$, $u_{0}=10^{-5}$, $\alpha_{g,0}=0.1$, and $\alpha_{e,0}=1$. When $\ell\rightarrow 10^5$, they approach to the fixed point value, $c^{*}=v^{*}=u^{*}=0.199$.}\label{fig:finite_c}
\end{figure}

In above three cases, we consider the velocity of Coulomb interaction is infinite, i.e., $c=\infty$. Here, we will consider the finite $c$. In this situation, the action is
\begin{align*}
S=&\int dx^{3}d\tau\;\left[\Psi^{\dagger}(\partial_{\tau}-iv\Gamma_{a}\partial_{a})\Psi+\frac{1}{2}(\partial_{\tau}\phi)^{2}/u^{2}+(\nabla\phi)^{2})+\frac{1}{4!}\frac{\lambda}{u}\phi^{4}+\phi(\Psi^{\dagger}M\Psi)\right]\\
&+\int dx^{3}d\tau\;\left\{\frac{1}{2}\left[ (\partial_{\tau}A_{a}/c+\nabla_{a}A_{0})^{2}-(\epsilon_{abc}\partial_{b}A_{c})^{2} \right]+ieA^{0}(\Psi^{\dagger}\Psi)+ie\frac{v}{c}A^{a}(\Psi^{\dagger}\Gamma_{a}\Psi)\right\}.
\end{align*}
where $\Psi$, $\phi$, and $A^{\mu}$ are for the fermion, order parameter, and photon ($a=x,y,z$).
So, we have three velocities, $u$, $v$, and $c$ which are for the order parameter, fermion, and the photon, respectively. Due to this, we have additional parameter, $x:=v/c$. The flow equations are given by
\begin{align*}
\frac{dx}{d\ell}=&-\frac{\alpha_{g}}{\pi}\frac{2xy(1-y)}{3(1+y)^{2}}+\frac{\alpha_{e}}{\pi}x\left( \frac{N_{f}}{3}(1+x)+\frac{2(1+3x+4x^{2})}{(1+x)^{2}} \right)(1-x),\\
\frac{dy}{d\ell}=&\frac{\alpha_{g}}{\pi}\left( \frac{N_{f}}{2}(1+y)+\frac{2y}{3(1+y)^{2}}  \right)(1-y)y-\frac{2\alpha_{e}}{3\pi}\frac{y(1-x)}{(1+x)^{2}}(1+3x+4x^{2}),\\
\frac{d\alpha_{g}}{d\ell}=&-\frac{\alpha_{g}^{2}}{\pi}\left( N_{f}+\frac{2y(5+4y)}{3(1+y)^{2}} \right)+\frac{2\alpha_{g}\alpha_{e}}{3\pi}\frac{(11+13x-x^{2}-5x^{3})}{(1+x)^{2}},\\
\frac{d\alpha_{e}}{d\ell}=&-\frac{2\alpha_{e}^{2}}{3\pi}N_{f}-\frac{2\alpha_{e}^{2}}{3\pi}\frac{(1-x)}{(1+x)^{2}}(1+3x+4x^{2})+\frac{2\alpha_{g}\alpha_{e}}{3\pi}\frac{y(1-y)}{(1+y)^{2}},\\
\frac{d\lambda}{d\ell}=&-\frac{3\lambda^{2}}{16\pi^{2}}+48N_{f}y\alpha_{g}^{2}-\frac{\lambda\alpha_{g}}{2\pi}N_{f}(3+y^{2}),
\end{align*}
and the fixed point is $(x^{*},y^{*},\alpha_{g}^{*},\alpha_{e}^{*},\lambda^{*})=(1,1,0,0,0)$. 
Note that for the initial value $c_{0}=1$, $v_{0}=10^{-2}$, $u_{0}=10^{-5}$, (because the orders of usual ratios between the velocities are $\mathcal{O}(v/c)\approx10^{-2}$ and $\mathcal{O}(u/c)\approx10^{-5}$) $\alpha_{g,0}=0.1$, and $\alpha_{e,0}=1$ (because of $\alpha_{e}\approx \alpha c/v\approx1$ where $\alpha\approx10^{-2}$ is the fine structure constant), to approach to the fixed point, we need to $\ell\rightarrow10^{5}$ (Fig.\ref{fig:finite_c}). 

At the fixed point, we have $v=u=c$, that is, the system become relativistic (so $x=y=1$). Near the fixed point, the flow equations for $\alpha_{e}$, $\alpha_{g}$, and $\lambda$ are
\begin{align*}
\frac{d\alpha_{e}}{d\ell}=&-\frac{2\alpha_{e}^{2}}{3\pi}N_{f},\\
\frac{d\alpha_{g}}{d\ell}=&-\frac{\alpha_{g}^{2}}{\pi}\left(N_{f}+\frac{3}{2}\right)+\frac{3\alpha_{e}\alpha_{g}}{\pi},\\
\frac{d\lambda}{d\ell}=&-\frac{3\lambda^{2}}{16\pi^{2}}+48N_{f}\alpha_{g}^{2}-\frac{2N_{f}}{\pi}\lambda\alpha_{g}.
\end{align*}
By the same way as before, $\mathcal{R}_{G}$, $\mathcal{R}_{\alpha}$, and $\mathcal{R}_{V}$ are
\begin{align*}
\mathcal{R}_{G,\text{rel}}=&\frac{\sqrt{2}}{3}\sqrt{\frac{-15N_{f}-2N_{f}^{2}+\sqrt{N_{f}(2916+1521N_{f}+204N_{f}^{2}+4N_{f}^{3})}}{9+2N_{f}}},\\
\mathcal{R}_{\alpha,\text{rel}}=&\frac{3(9+2N_{f})}{2(3+2N_{f})},\\
\mathcal{R}_{V,\text{rel}}=&1.
\end{align*}

\begin{figure}[h]
\centering
\subfigure[]{
\includegraphics[width=0.322\linewidth]{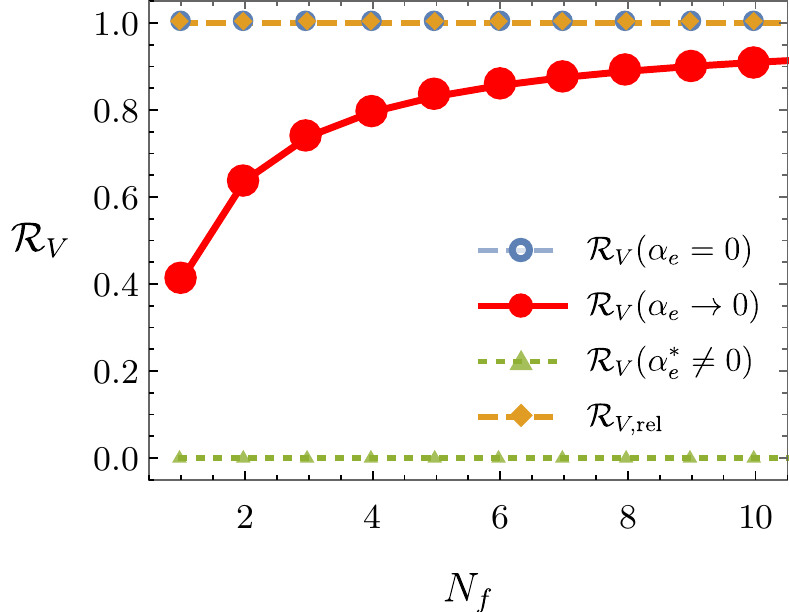}\label{fig:velocity_ratio}}
\subfigure[]{
\includegraphics[width=0.322\linewidth]{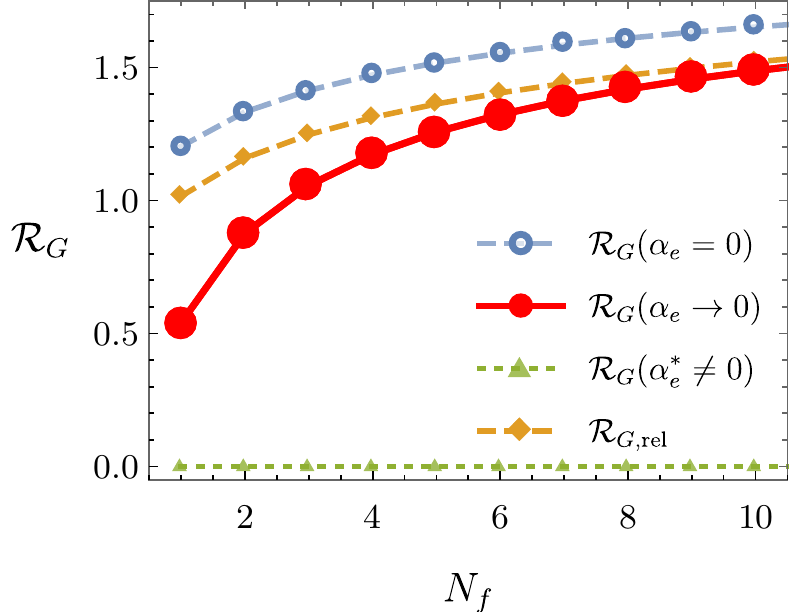}\label{fig:gap_ratio}}
\subfigure[]{
\includegraphics[width=0.322\linewidth]{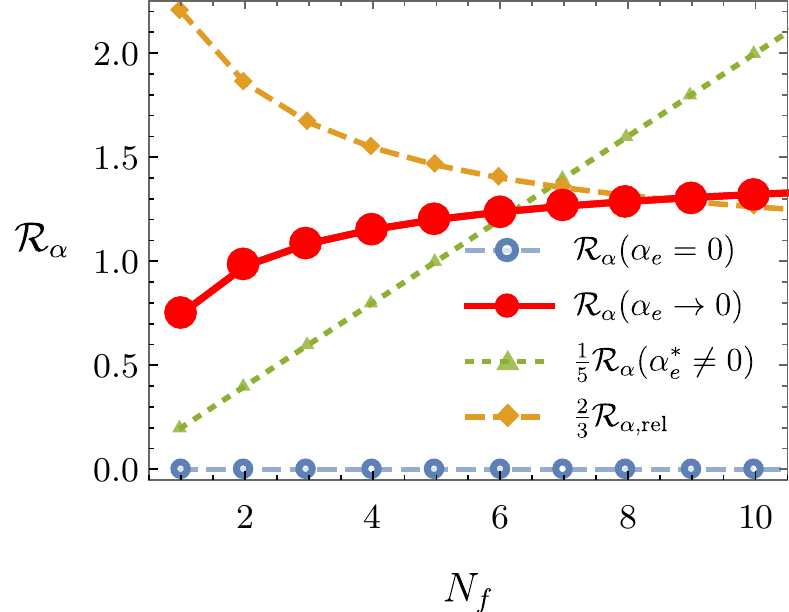}\label{fig:coupling_ratio}}

\caption{$N_{f}$ dependence of $\mathcal{R}_{V}$, $\mathcal{R}_{G}$, and $\mathcal{R}_{\alpha}$ for $\alpha_{e}=0$, $\alpha_{e}\rightarrow0$, $\alpha_{e}^{*}\neq0$, and relativistic regime. (a), (b), and (c) are for the velocity ratio, gap ratio, and coupling ratio, respectively. In (a), $\mathcal{R}_{V}(\alpha_{e}=0)$ and $\mathcal{R}_{V,\text{rel}}$ are equal to 1, and $\mathcal{R}_{V}(\alpha_{e}\rightarrow0)$ goes to 1 as $N_{f}\rightarrow\infty$. $\mathcal{R}_{V}(\alpha_{e}^{*}\neq0)$ is always 0. In (b), $\mathcal{R}_{G}(\alpha_{e}=0)$, $\mathcal{R}_{G}(\alpha_{e}\rightarrow0)$, and $\mathcal{R}_{G}(\alpha_{e}^{*}\neq0)$ approach to 2. In (c), $\mathcal{R}_{\alpha}(\alpha_{e}^{*}\neq0)$ and $\mathcal{R}_{\alpha,\text{rel}}$ are scaled by $1/5$ and $2/3$, respectively. $\mathcal{R}_{\alpha}(\alpha_{e}\rightarrow0)$ and $\mathcal{R}_{\alpha,\text{rel}}$ approach to $3/2$, but $\mathcal{R}_{\alpha}(\alpha_{e}^{*}\neq0)$ is linear in $N_{f}$. $\mathcal{R}_{\alpha}(\alpha_{e}=0)=0$ because $\alpha_{e}=0$.  }\label{fig:univ_ratio}
\end{figure}

The summaries for all the limits are presented at Fig.\ref{fig:univ_ratio}.

\section{Disccusions for general anisotropic case}
Let us discuss for the general anisotropic case, which is for $v_{x}\neq v_{y}\neq v_{z}$.\\
Let $v_{i}$, $u_{i}$ and $c_{i}$ be the velocities of $i$-direction for fermion, order parameter, and Coulomb interaction. In general anisotropic case, $v_{x}\neq v_{y}\neq v_{z}$, $u_{x}\neq u_{y} \neq u_{z}$, and $c_{x}\neq c_{y}\neq c_{z}$, near the fixed point, we have the following relations,
\begin{align*}
&\frac{u_{x}}{u_{y}}=\frac{v_{x}}{v_{y}}=\frac{c_{x}}{c_{y}},\\
&\frac{u_{x}}{u_{z}}=\frac{v_{x}}{v_{z}}=\frac{c_{x}}{c_{z}},\\
&\frac{u_{y}}{u_{z}}=\frac{v_{y}}{v_{z}}=\frac{c_{y}}{c_{z}}.\\
\end{align*}
These mean that we only need $u\equiv u_{x}$, $v\equiv v_{x}$, and $s\equiv c_{x}$ because all the velocity for the different directions are connected by some ratio.\\
The flow equations are same as isotropic case with divided by $v_{y}/v_{x}$. They has the same fixed point with isotropic case.

\section{With $\text{SO}(N_{b})$  symmetric scalar theory}
Let us consider the extended model which $N_{f}$ Dirac fermions are coupled with the instantaneous Coulomb interaction and $\text{SO}(N_{b})$ symmetric $N_{b}$ scalar bosons. In this case, we have the following actions,
\begin{align*}
S_{\text{tot}}=&S_{\Psi}+S_{\phi}+S_{\varphi}+S_{\Psi\phi}+S_{\Psi\varphi},\\
S_{\Psi}=&\int d^{3}xd\tau\;\Psi^{\dagger}( \partial_{\tau}-i v\gamma_{\mu}\partial_{\mu})\Psi,\\
S_{\phi}=&\int d^{3}xd\tau\;\left[ \frac{1}{2}(\partial_{\tau}\vec{\phi})^{2}/u^{2}+\frac{1}{2}(\nabla\vec{\phi})^{2}+\frac{1}{2}r\vec{\phi}^{2}+\frac{1}{4!}\frac{\lambda}{u}(\vec{\phi}^{2})^{2} \right],\\
S_{\varphi}=&\int d^{3}xd\tau\;\frac{1}{2}(\nabla\varphi)^{2},\\
S_{\Psi\phi}=&g\int d^{3}xd\tau\;\vec{\phi}(\Psi^{\dagger}\vec{M}\Psi),\\
S_{\Psi\varphi}=&ie\int d^{3}d\tau\;\varphi(\Psi^{\dagger}\Psi),
\end{align*}
where $\vec{\phi}=(\phi_{1},\phi_{2},\cdots,\phi_{N_{b}})$ is in the fundamental representation of $\text{SO}(N_{b})$, $\vec{M}=(M_{1},M_{2},\cdots,M_{N_{b}})$ is also in the fundamental representation of $\text{SO}(N_{b})$. Here, $\gamma_{\mu}$ ($\mu=1,2,3$) and $M_{i}$ ($i=1,2,\cdots,N_{b}$)  are extended Gamma matrices and defined by
\begin{align*}
\gamma_{\mu}:=&\Gamma_{\mu}\otimes I_{2^{\lfloor N_{b}/2 \rfloor}},\;(\mu=1,2,3)\\
M_{i}:=&M\otimes \bar{\gamma}_{i},\;(i=1,2,\cdots,N_{b})
\end{align*}
where $I_{N}$ is $N$-dimensional identity matrix and $\Gamma_{\mu}$'s are $4N_{f}\times 4N_{f}$ matrices which are introduced in main text. $\lfloor x\rfloor$ means the floor function which is defined by $\lfloor x\rfloor=\text{max}\{m\in\mathbb{Z}|m\leq x\}$. $\gamma_{\mu}$ and $M_{i}$ satisfy that $\{\gamma_{\mu},\gamma_{\nu}\}=2\delta_{\mu\nu}I_{4N_{f}}\otimes I_{2^{\lfloor N_{b}/2 \rfloor}}$, $\{M_{i},M_{j}\}=2\delta_{ij}I_{4N_{f}}\otimes I_{2^{\lfloor N_{b}/2 \rfloor}}$, and $\{\gamma_{\mu},M_{j}\}=0$. $\bar{\gamma}_{i}$'s are $2^{\lfloor N_{b}/2\rfloor}$ dimensional matrices which satisfy the $N_{b}$ dimensional Clifford algebra, $\{\bar{\gamma}_{i},\bar{\gamma}_{j}\}=2\delta_{ij}I_{2^{\lfloor N_{b}/2\rfloor}}$.

The flow equations are
\begin{align*}
\frac{dy}{d\ell}=&\frac{\alpha_{g}}{\pi}\left( \frac{N_{f}A(N_{b})}{2}(1+y)+\frac{2N_{b}}{3}\frac{y}{(1+y)^{2}} \right)(1-y)y-\frac{2\alpha_{e}}{3\pi}y,\\
\frac{d\alpha_{g}}{d\ell}=&-\frac{\alpha_{g}^{2}}{\pi}\left( N_{f}A(N_{b})+\frac{2}{3}\frac{N_{b}y(5+4y)}{(1+y)^{2}} \right)+\frac{4\alpha_{g}\alpha_{e}}{3\pi},\\
\frac{d\alpha_{e}}{d\ell}=&-\frac{2\alpha_{e}^{2}}{3\pi}(1+N_{f}A(N_{b}))+\frac{2\alpha_{g}\alpha_{e}}{3\pi}\frac{N_{b}y(1-y)}{(1+y)^{2}},\\
\frac{d\lambda}{d\ell}=&-\frac{3\lambda^{2}}{16\pi^{2}}\frac{(N_{b}+8)}{9}+48N_{f}A(N_{b})y\alpha_{g}^{2}-\frac{N_{f}A(N_{b})}{2}\alpha_{g}\lambda(3+y^{2}).
\end{align*}
where $A(N_{b})\equiv2^{\lfloor N_{b}/2 \rfloor}$. Here, $y\equiv u/v$ is  less than 1 for finite $N_{f}$ and $N_{b}$, so the fermion velocity is still faster than the boson velocity. Similar to the original model, we can calculate the universal ratios in terms of $N_{f}$ and $N_{b}$. For example, when $N_{b}=1$, $A(N_{b}=1)=1$, so we can recover the results in main text. \\

\section{$d=3-\epsilon$ calculation}
Our calculations can be straightforwardly generalized to ones in $d=3-\epsilon$ dimensions. 
The action becomes
\begin{align*}
\mathcal{S}=&\int d^dx d\tau \sum_{a=1}^{N_{f}}\left[ \psi_{a}^{\dagger}(\partial_{\tau}+\mathcal{H}(-i\nabla))\psi_{a}+ie\varphi(\psi_{a}^{\dagger}\psi_{a})+g\phi(\psi_{a}^{\dagger}M\psi_{a})\right]\\
&+ \int d^dx d\tau \left[\frac{1}{2}|\nabla|^{d-1}(\varphi)^{2}+\frac{1}{2}\left(\frac{(\partial_{\tau}\phi)^{2}}{u}+(\nabla\phi)^{2}\right)+\frac{1}{4!}\frac{\lambda}{u}\phi^{4}\right].
\end{align*}
Notice  that the Coulomb potential $\varphi$ has the non-analytic propagator below $d=3$. This is because the Coulomb interaction is $1/r$ potential in any spatial dimensions. 
With this set-up, we can investigate the long-range Coulomb interactions at the chiral symmetry breaking systematically. 

The scaling dimensions of the coupling constants at the tree level are   
\begin{eqnarray}
[\alpha_g]=\epsilon, \quad [\alpha_e]=0, \quad [\lambda]=\epsilon, \quad [y]=0. \nonumber
\end{eqnarray}
Then it is straightforward to do $\epsilon$-expansion and we obtain the RG equations as follows. 
\begin{align*}
\frac{dy}{d\ell}=&y\left[\frac{\alpha_{g}}{\pi}\left( \frac{N_{f}}{2}(1-y^{2})+\frac{2y(1-y)}{3(1+y)^{2}} \right)-\frac{2}{3}\frac{\alpha_{e}}{\pi}\right],\\
\frac{d\lambda}{d\ell}=&\epsilon\lambda-\frac{3\lambda^{2}}{16\pi^{2}}+48N_{f}y\alpha_{g}^{2}-\frac{\alpha_{g}}{2\pi}N_{f}\lambda(3+y^{2}),\\
\frac{d\alpha_{g}}{d\ell}=&\epsilon\alpha_{g}-\frac{\alpha_{g}^{2}}{\pi}\left[ N_{f}+\frac{2y(5+4y)}{3(1+y)^{2}} \right]+\frac{4}{3}\frac{\alpha_{e}\alpha_{g}}{\pi},\\
\frac{d\alpha_{e}}{d\ell}=&\frac{\alpha_{g}\alpha_{e}}{\pi}\frac{2y(1-y)}{3(1+y)^{2}}-\frac{\alpha_{e}^{2}}{\pi}\frac{2}{3}(1+N_{f}).
\end{align*}
 
Striking differences in RG flow appears in the lower dimension ($\epsilon \neq0$). 
We find one stable fixed point, \begin{align*}
(y^{*},\lambda^{*},\alpha_{g}^{*},\alpha_{e}^{*})=\left(1,\frac{8\pi^{2}\epsilon}{3(3+2N_{f})}(3-2N_{f}+\sqrt{9+132N_{f}+4N_{f}^{2}}),\frac{2\pi\epsilon}{3+2N_{f}},0\right),
\end{align*}
no matter how we set $\alpha_e (l=0) \neq 0$ or $\alpha_e (l=0) = 0$.
Notice that the velocity ratio becomes unity $y^*=1$, so the Lorentz invariance is emergent in sharp contrast to the case in $d=3$ as discussed in the main-text.  
This is because the quasi-particles are strongly renormalized by the Yukawa interaction and the self-interaction, so the quasi-particles become ill-defined. Then, the long range Coulomb interaction becomes sub-dominant.

\bibliographystyle{apsrev4-1}

\end{document}